\documentclass{ws-ijmpa}
\usepackage[super,compress]{cite}
\usepackage{graphicx}

\newcommand{\pipipipi}{\mbox{$\pi^+\pi^-\pi^+\pi^-$ }}

\newcommand{\pipi}{\mbox{$\pi^{+}\pi^{-}$} }
\newcommand{\kk}{\mbox{$K^{+}K^{-}$} }

\newcommand{\pom}{$I\hspace{-1.6mm}P$}

\begin{document}
\markboth{Andrew Kirk}
{A review of central production experiments at the CERN Omega spectrometer}

%
\catchline{}{}{}{}{}
%

\title{
A REVIEW OF CENTRAL PRODUCTION EXPERIMENTS AT THE CERN OMEGA SPECTROMETER
}
\author{ANDREW KIRK}

\address{
Culham Centre for Fusion Energy, Abingdon, Oxfordshire, United Kingdom \\
andrew.kirk@ccfe.ac.uk}

\maketitle
\begin{history}
\received{}
\revised{}
\end{history}

\begin{abstract}
The non-Abelian nature of QCD suggests that
particles that have a gluon constituent, such as glueballs or
hybrids, should exist. 
This paper presents a study of central meson production in the fixed target experiments WA76, WA91 and WA102 at the CERN Omega spectrometer at
centre-of-mass energies of
$\sqrt{s} = 12.7$, 23.8 and 29~GeV.
A study of the resonance production cross section as a function of $\sqrt{s}$ shows which states are compatible 
with being produced by Double Pomeron Exchange (DPE).
In these DPE processes, the difference in the transverse momentum between the exchange particles ($dP_T)$ can be used to select out known $q\overline q$ 
states from non-$q \overline q$ candidates.  The distribution of the azimuthal angle ($\phi$) between the two exchange particles suggests that the 
Pomeron transforms like a non-conserved vector current.
Finally there is evidence from an analysis of the the decay modes of the scalar states 
observed, that the lightest scalar glueball manifests itself through the
mixing with nearby $q\overline q$ states.
\end{abstract}

\ccode{PACS numbers:12.39.Mk, 14.40.Be, 11.55.Jy, 13.85.Ni}


\section{Introduction}

Quantum ChromoDynamics (QCD)
not only describes how quarks and antiquarks interact, but also
predicts that the gluons, which are the quanta of the field, will themselves
interact to form mesons.
If the object formed is composed entirely of valence gluons the meson
is called a glueball, however if it is composed of a mixture of
valence quarks, antiquarks
and gluons (i.e. $ q \overline q g$ ) it is called a hybrid.
In addition, $ q \overline q q \overline q $ states are also predicted.
\par               
The best estimate for the masses of glueballs comes from 
lattice gauge theory calculations
\cite{re:lgt}
which show that the lightest glueball has $J^{PC}$~=~$0^{++}$ and that
\begin{center}
$m(2^{++})/m(0^{++}) \approx 1.5 $
\end{center}
and  depending on how the lattice parameters are extrapolated to the
mass scale that
\begin{center}
$m(0^{++}) =(1500-1750) $ MeV.
\end{center}
The mass of the $0^{-+}$ glueball is predicted to be similar to that of the
$2^{++}$ glueball whilst glueballs with other quantum numbers
are predicted to be higher in mass.
\par
The flux tube model has been used to calculate the masses of the lowest
lying hybrid states and predicts 
\cite{re:ISGURBNL}
that
\begin{center}
$m(1^{--},0^{-+},1^{-+},2^{-+}) \approx 1900$~MeV.
\end{center}
\par
Since the lightest non-$q \overline q$ states
are predicted to have the same quantum numbers and lie in the same mass region as $q \overline q$ states 
one way to find them is 
to look for extra states, that is states that have quantum numbers
of already completed nonets and that have masses which are sufficiently
low that they are unlikely to be members of the radially excited nonets.
It was hoped that these extra states would have unusual branching ratios and/or be preferentially produced in gluon rich processes
such as Double Pomeron Exchange (DPE), where  
the Pomeron trajectory is 
thought to be
mediated by the exchange of a virtual multi-gluon state.
\par
The CERN fixed target experiments WA76, WA91 and WA102, which were performed at the Omega spectrometer were 
designed to study exclusive final states
formed in the reaction
\begin{center}
$pp$ $\longrightarrow$p$_{f}X^{0}$p$_s$,
\end{center}
where the subscripts $f$ and $s$ refer to the fastest and slowest
particles, identified as protons, in the laboratory frame respectively and $X^0$ represents
the central system. Such reactions are expected to
be mediated by double exchange processes
where both Pomeron and Reggeon exchange can occur.
\begin{figure}[ht]
\centerline{\includegraphics[width=12cm]{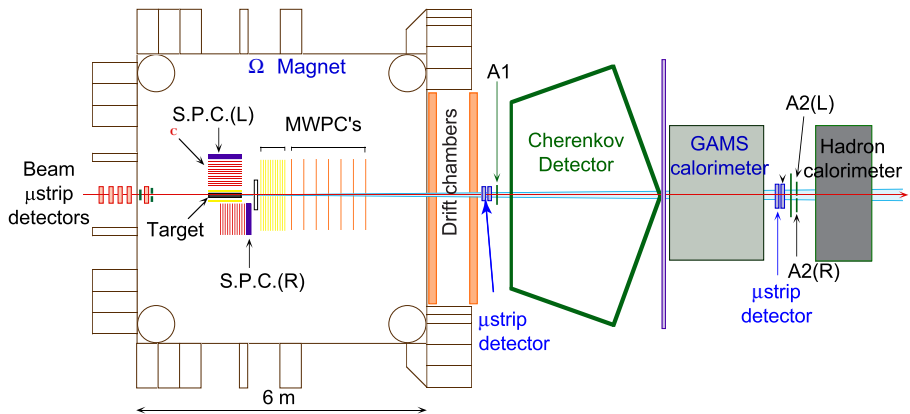}}
\caption{Layout of the $\Omega$ spectrometer for the 1996 run of experiment WA102.}
\label{fi:omega}
\end{figure}
\par
The CERN NA12/2 experiment 
contributed to this programme through the study of neutral decay modes of light mesons using the GAMS 4000 electromagnetic calorimeter \cite{re:GAMS}.
The WA76 and WA91 experiments, which were performed at the Omega spectrometer concentrated 
mainly on the decay to charged particles using proton beams at 85, 300 and 450 GeV corresponding 
to
centre-of-mass energies of
$\sqrt{s} = 12.7$, 23.8 and 29~GeV.
In 1995 and 1996 experiment WA102 combined the excellent charged particle reconstruction from the 
CERN Omega spectrometer with the neutral particle identification offered by GAMS 4000.
The main results shown in this paper are from the WA102 experiment, 
which searched for
non-$q \overline q$ mesons in the mass range up to 2.5~GeV.
The layout of 
the 1996 run of WA102 is
shown in fig.~\ref{fi:omega}.
The positively charged H1 beam in the west area was incident on a
60~cm long liquid hydrogen target.
A set of 
ten 20~$\mu$m pitch micro-strip
detectors were used
to perform an accurate measurement of the incident beam's trajectory.
The outgoing fast track, which had a momentum in the range 300 to 450~GeV,
was measured by two sets of 4 micro-strip detectors
placed 8 and 12~m downstream from the target.
The Omega Multi Wire Proportional Chambers (MWPCs) and Drift Chambers
were used to measure the medium momentum
tracks ($\approx$~1 to 40~GeV)
leaving the interaction region, with particle identification coming from 
a threshold Cerenkov counter (C1).
Photon detection was provided by the
GAMS-4000 lead glass calorimeter.
The trigger was designed to enhance double exchange
processes with respect to single exchange and elastic processes.
Details of the trigger conditions, the data
processing and event selection 
have been given in previous publications~\cite{re:expt}.
\par 
In this paper the  status of these experiments is reviewed. In section 2 the possibility
that a kinematic filter exists in central production that can discriminate between gluonic and $q \overline q$ states is discussed 
while in section 3 and 4 the status of the search for the
scalar and tensor glueball is presented.

\section{The effect of kinematic variables on central meson production}
The experiments have been 
performed at incident beam momenta of 85, 300 and 450 GeV, corresponding to
centre-of-mass energies of
$\sqrt{s} = 12.7$, 23.8 and 29~GeV.
Theoretical 
predictions \cite{pred} of the evolution of 
the different exchange mechanisms with centre
of mass energy, suggest that 
\begin{center}
$\sigma$(RR) $\sim s^{-1}$,\\
$\sigma$(R\pom) $\sim s^{-0.5}$,\\
$\sigma$(\pom\pom) $\sim$ constant,
\end{center}
where RR, R\pom \thinspace \thinspace 
and \pom\pom \thinspace \thinspace refer
to Reggeon-Reggeon, Reggeon-Pomeron and Pomeron-Pomeron
exchange respectively. Hence it is expected that Double Pomeron Exchange
(DPE) would become more significant at high energies, whereas the Reggeon-Reggeon and 
Reggeon-Pomeron mechanisms would decrease in importance. 
\begin{figure}[ht]
\centerline{\includegraphics[width=12cm]{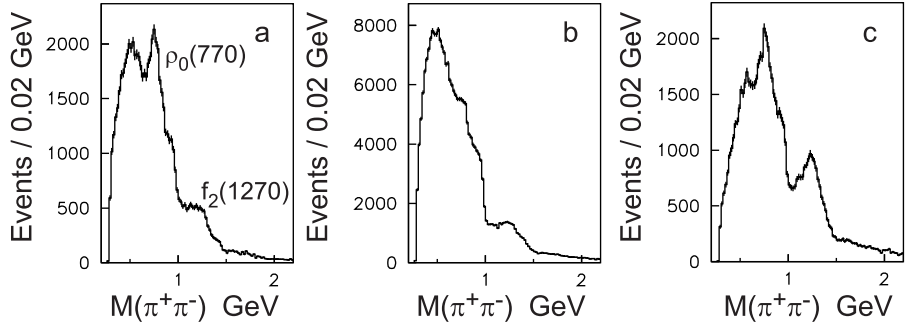}}
\caption{
The centrally produced \pipi effective mass spectrum at 
$\sqrt{s}$  = a) 12.7 GeV and b) 29 GeV using a LL trigger and c) at 29 GeV from a LR trigger.
}
\label{fi:pipiofs}
\end{figure}
\par
The decrease of the non-DPE cross section with energy can be inferred
by comparing the \pipi mass spectrum obtained in $pp$ interactions, 
under the same trigger conditions, from WA76 at $\sqrt{s}$  = 12.7 GeV 
(fig.~\ref{fi:pipiofs}a) 
and from WA102 at $\sqrt{s}$  = 29 GeV (fig.~\ref{fi:pipiofs}b).  
The \pipi mass spectra for the two cases show that     
the signal-to-background ratio for the $\rho^0$(770)
is much lower at high energy.  The WA76 collaboration reported
that the ratio of the $\rho^0$(770) cross sections at 23.8 GeV and 12.7 GeV
is 0.44~$\pm$~0.07  \cite{wa76}.
Since isospin 1 states such as the $\rho^0$(770) cannot be produced by DPE, 
the decrease 
of the $\rho^{0}(770)$ signal at high $\sqrt{s}$
is consistent with DPE becoming 
relatively more important with increasing energy with respect to other
exchange processes.
Due to the fact that there was no electromagnetic calorimeter in 
the WA76 experiment at $\sqrt{s}$ of 12.7 GeV, the $s$ dependence of
only a subset of states could be determined.
The states that were compatible with being produced by DPE were: $\eta^\prime$,
$f_0(980)$, $f_0(1500)$, $f_1(1285)$, $f_1(1420)$, $f_2(1270)$ and $f_2(1950)$ \cite{sumpap}.
\par
A comparison of the production cross-section for
all the resonances
observed in the WA102 experiment at $\sqrt s$~=~29.1~GeV
\cite{sumpap} shows that 
$s \bar{s}$ states are produced much more weakly than
$n \bar{n}$ states (i.e. those containing u and d quarks). 
For example,     
the cross section for the production of the $f_2(1270)$, whose production
has been found to be consistent with DPE, is
more than 40 times greater than the cross section of the $f_2^\prime(1525)$.
Part of this suppression could be due to the fact that there is a 
kinematic suppression of $M_{X^0}^{-2}$
and some could be explained by the fact that the centre of mass energy
dependence of the $f_2(1270)$ allows for up to a 30~\% contribution
from non-DPE processes. However, a suppression of $\approx~20$ still
needs to be explained.
Hence there could be some strong dependence on 
the mass of the produced quarks in DPE.
\begin{figure}[ht]
\centerline{\includegraphics[width=12cm]{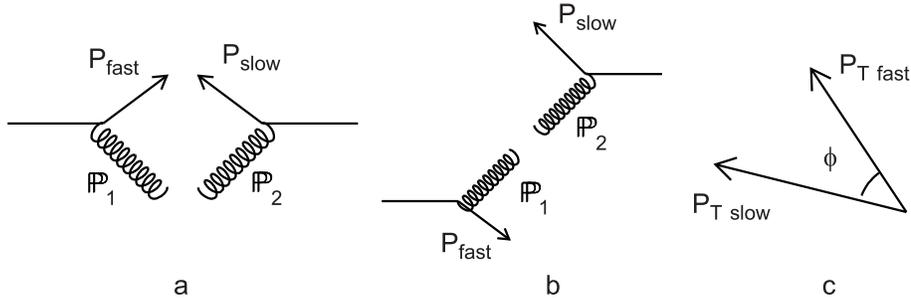}}
\caption{Schematic diagrams 
in the centre of mass for a) LL and b) LR triggers. 
c) Definition of the azimuthal angle $\phi$ between the $P_T$ vectors 
of the outgoing protons}
\label{fi:dpe}
\end{figure}
\par
The fact that known $q\bar{q}$ states are also seen in DPE initially frustrated
the hope that such experiments would prove to be a clean glueball
source. However, during the experiments an additional kinematic variable was found, which was initially exposed due to a change
in the trigger conditions that were implemented moving from WA76 to WA91. 
In the WA76 experiment in order to remove the much more frequently occurring $pp$ 
elastic scattering events the trigger required that the two outgoing protons were on the
same side relative to the incident beam (classified as LL) and shown pictorially in
fig. \ref{fi:dpe}a.  In WA91 a second trigger condition was implemented,
which allowed the detection of the decay of the centrally produced meson to charged 
particles. This allowed configurations to be recorded where the outgoing protons were on 
opposite sides of the beam (classified as LR), which is shown pictorially 
in fig. \ref{fi:dpe}b. 
The azimuthal angle $\phi$ which is defined as the angle between the $p_T$ 
vectors of the two outgoing protons (see fig. \ref{fi:dpe}c) is small 
for the LL configuration and near to 180 degrees for the LR configuration. 
The WA91 collaboration found that 
there was a difference in 
the resonances observed for these two trigger configurations~\cite{wa91corr}. 
Fig. \ref{fi:pipiofs}b and c compares the \pipi mass spectrum obtained with  
the LL and LR trigger configuration from WA102 at $\sqrt{s} = $29~GeV. 
In the LR configuration, the $\rho^0$(770) and $f_2(1270)$, known $q \overline q$ states  
are very prominent. The SFM collaboration at the ISR had also observed that the directions 
taken by the outgoing fastest and slowest particles were correlated with the production of 
the $f_2(1270)$ \cite{ref:SFM}, however, in that case they concluded that the observation 
may suggest that the $f_2(1270)$ may have a large glue content. In the case of 
WA91, since the $\rho^0(770)$ and $f_2(1270)$ have a similar dependence
the conclusion drawn was that the  configuration of the outgoing protons 
 may provide a way of distinguishing between $q \overline q$ and 
non-$q \overline q$ states.  
\par
 In fact, the WA91 collaboration \cite{wa91corr}
showed that there 
were a  wide variety of resonances that favoured production 
when the angle between the outgoing slow and fast protons was near to 0 degrees
compared to when the angle was near to 180 degrees.
In order to try to explain this effect
in terms of a physical model,
Close and Kirk~\cite{closeak}
proposed that the data be analysed
in terms of the parameter $dP_T$, which is the
difference in transverse momentum
between the particles exchanged from the
fast and slow vertices.
\par
The WA102 collaboration presented studies of how different
resonances were produced as a function of the 
parameter $dP_T$~(see \cite{sumpap} and references therein).                            
The fraction of each resonance 
was calculated for
$dP_T$$\leq$0.2 GeV  and $dP_T$$\geq$0.5 GeV and the ratio of the
production at small $dP_T$ to large $dP_T$ was found to be a useful discriminant \cite{sumpap}. 
It was found that all the undisputed $q \overline q$ states 
(for example, $\eta$, $\eta^{'}$, $f_1(1285)$, 
$f_1(1420)$, $f_2(1270)$ and $f_2(1525)$) 
which can be produced in DPE, namely those with positive G parity and $I=0$,
have a very small value for this ratio ($\leq 0.1$).
However, the
interesting states, which could have a non-$q \overline q $ 
or gluonic component have 
a large value for this ratio~\cite{sumpap}, 
for example $f_0(980)$, $f_0(1500)$, $f_0(1710)$, $f_2(1910)$ and $f_2(1950)$.
\begin{figure}[ht]
\centerline{\includegraphics[width=12cm]{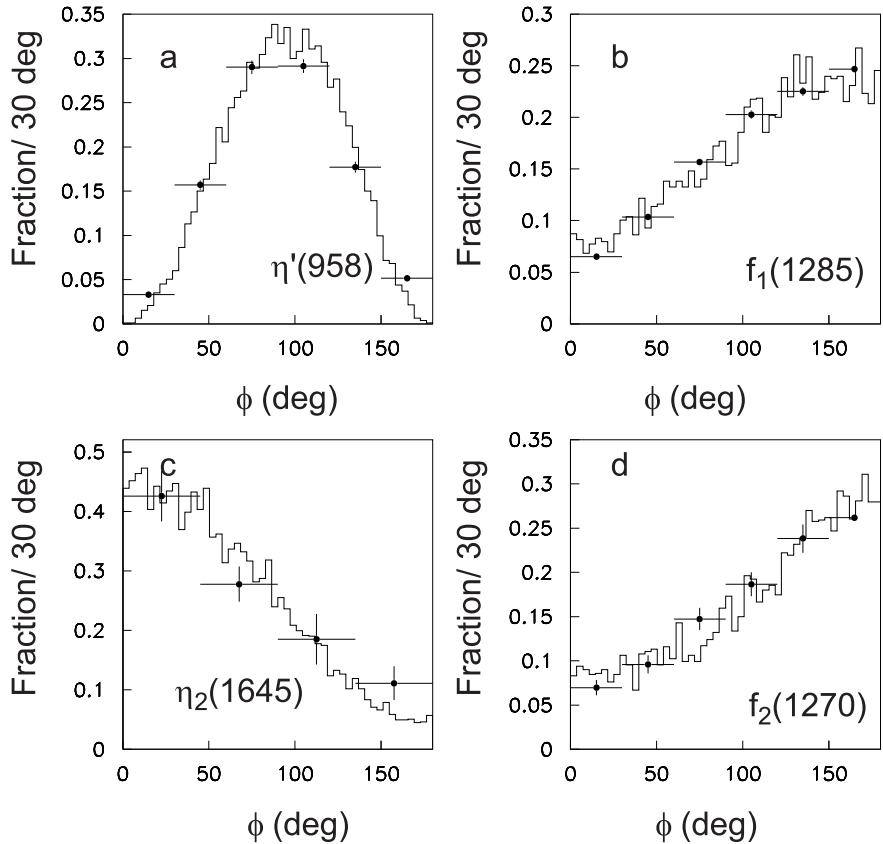}}
\caption{The $\phi$ dependence for undisputed $q \overline q$ states a) $\eta^\prime$, b) $f_1(1285)$, c) $\eta_2(1645)$ and d) $f_2(1270)$ for the data (dots) and the model (histogram).
}
\label{fi:phiqqbar}
\end{figure}
\par
In addition to the $dP_T$ dependencies, 
an interesting effect was observed in
the azimuthal angle $\phi$ which is defined as the angle between the $p_T$ 
vectors of the two outgoing protons (see \ref{fi:dpe}c).
Previously it had been assumed that the Pomeron, which is said to have ``vacuum quantum
numbers", transforms as a scalar and hence 
that the $\phi$ distribution 
would be flat for resonances
produced by DPE.
Fig. \ref{fi:phiqqbar} shows the $\phi$ dependence for the production of undisputed 
$q \overline q$ states with a range of spin quantum numbers, $J^{PC}$. 
The observed $\phi$ dependencies 
are clearly not flat and considerable variation
is observed among the resonances produced.
For the $q \overline q$ mesons that can be produced by DPE
the $\phi$ distributions 
maximise around $90^{o}$
for resonances with $J^{PC}$~=~$0^{-+}$, 
at $180^o$ for those with 
$J^{PC}$~=~$1^{++}$ and 
at $0^o$ 
for those with
$J^{PC}$~=~$2^{-+}$.       
\par
Several theoretical papers have been published on these 
effects~\cite{angdist,clschul}.
All agree that the exchanged particle
must have J~$>$~0
and that J~=~1
is the simplest explanation for the observed $\phi$ distributions.
Close and Schuler~\cite{clschul} have calculated the $\phi$ dependencies 
for the production of resonances with different $J^{PC}$
for the case where the exchanged particle is a 
Pomeron that transforms like a non-conserved
vector current.
In order 
to gain insight into the nature of the particles
exchanged in central $pp$ interactions 
Close, Kirk and Schuler~\cite{galuga} compared 
the predictions of this model with the data
for resonances with different $J^{PC}$ observed in the
WA102 experiment.
They found that for the production of $J^{PC}$~=~$0^{-+}$ mesons
they could predict the $\phi$ dependence as well as the observed 
vanishing cross section as $t \rightarrow 0$ absolutely.
Superimposed on fig. \ref{fi:phiqqbar}a is the prediction of this model. 
The model was also able to predict the
the $\phi$ dependence for the  
$J^{PC}$~=~$1^{++}$ and $2^{-+}$ mesons (fig. \ref{fi:phiqqbar}b and c) \cite{galuga}.
In the $0^{++}$ and $2^{++}$ sector the $\phi$ distributions
can be fitted with one parameter ($\mu^2$), where the sign of this parameter determines if the 
$\phi$ distribution is peaked at $0^o$ or $180^o$. For example, fig. \ref{fi:phiqqbar}d shows the $\phi$ 
dependence for the $f_2(1270)$ which can be well described with $\mu^2$ = -0.4 $GeV^2$. 
Understanding the dynamical origin of this sign then became a central issue
in the quest to distinguish $q \overline q$ states from
glueballs or other exotic states and this will be discussed in the following sections. 

\par
One of the first applications of these kinematic filters was to the axial vector nonet. 
At the start of the WA102 experiment the $f_1(1420)$, which was 
prominently observed in central production, was thought to be a non-$q \overline q$ candidate 
(see \cite{killf1} and references therein). There appeared to be three candidates
for the I=0 members of the $J^{PC} = 1^{++}$ nonet: $f_1(1285)$, $f_1(1420)$ and $f_1(1510)$.
However, the application of these kinematic selections 
showed that the $f_1(1285)$ and $f_1(1420)$
have the same behaviour; namely consistent with the
$f_1(1420)$ being the partner to the
$f_1(1285)$ in the $^3P_1$ nonet of axial mesons.
The conclusion reached from the analysis presented in ref. \cite{killf1} was 
that without confirmation of the existence of the
$f_1(1510)$ the isoscalar members of the $J^{PC}$~=~$1^{++}$ nonet
should be considered to be the $f_1(1285)$ and $f_1(1420)$ with a singlet-octet
mixing angle of approximately 50$^0$.

\section{The search for the scalar glueball}
In the 1990s the search for non-$q \overline q$ states became
possible due to the advent of high statistics experiments. 
To identify a scalar non-$q \overline q$ state
the first question to ask is how many $0^{++}$ states are there in the 
1-2 GeV mass region.
The problem was that in the scalar sector there appeared to be a 
multitude of states, with different experiments claiming new scalar states that had effectively no overlap in 
their mass and width.  
It quickly  became apparent that 
in order to compare results from different experiments                          
a unified and unitarised method of analysis was required.
\begin{figure}[ht]
\centerline{\includegraphics[width=12cm]{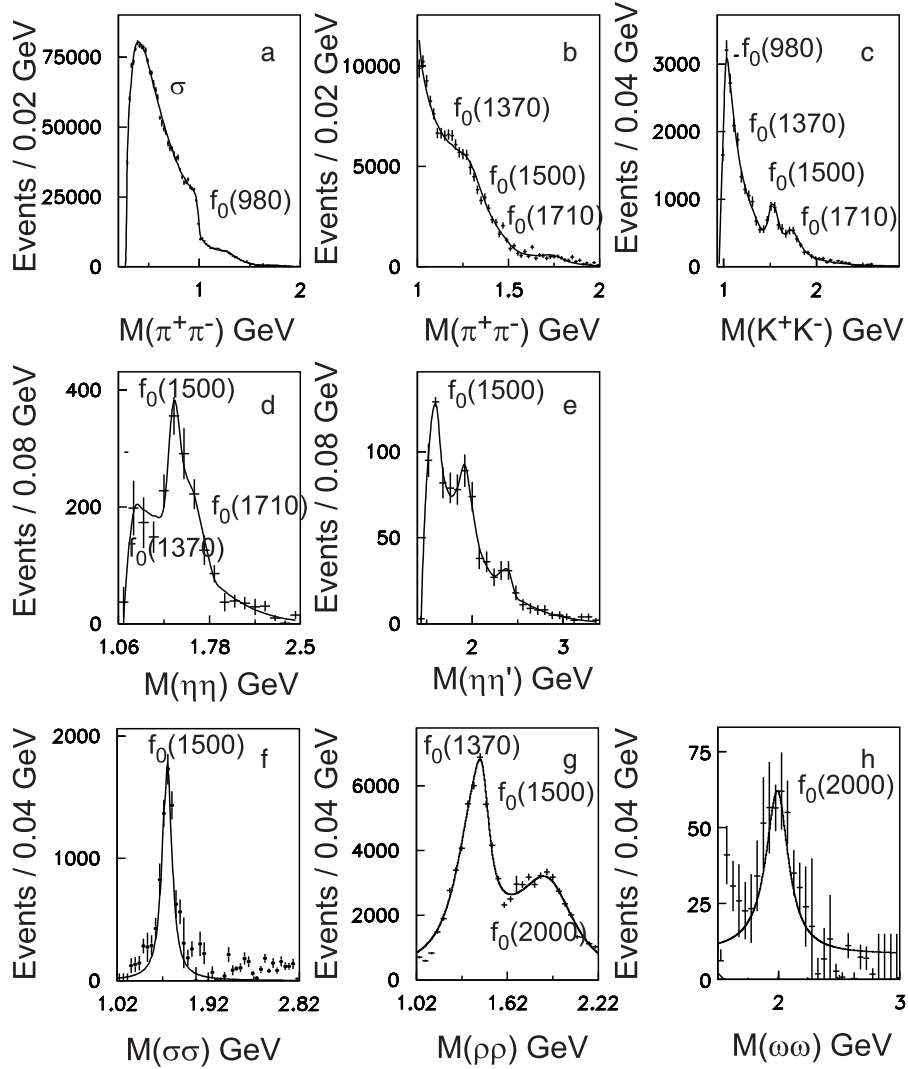}}
\caption{The S-wave contributions to the a),b) \pipi, c) \kk, d) $\eta\eta$, f) $\sigma\sigma$, g) $\rho\rho$ and h) $\omega\omega$ mass spectra 
and e) the total $\eta\eta^{'}$ mass spectra.
The location of the $J^{PC}$ = $0^{++}$ states identified are indicated.
}
\label{fi:swave}
\end{figure}
\par
A good example of why such an approach is required                            
can be seen by looking at what was being reported in the 1.5 GeV                                           
mass region in 1994.
In addition to the $f_0(1370)$ and $f_0(1710)$,                                                                
several experiments had observed apparently different                          
scalar states.                                                                  
The GAMS collaboration observed a G(1590) decaying to                                    
$\eta \eta $ and $\eta \eta^{\prime}$ \cite{re:GAMS}.                           
Experiments WA76 and WA91 observed a narrow state                               
($\Gamma \approx 50$ MeV) called the X(1450)                                    
in the centrally produced $\pi^+ \pi^- \pi^+ \pi^-$                             
channel \cite{re:wa914pi}.                                                      
While the Crystal Barrel experiment observed the $f_0(1500)$ in                   
several final states in $p \overline p $ annihilations \cite{re:cb}.            
Therefore there appeared to be three new scalar states around 1.5 GeV, each observed             
in different experiments.                                                       
However under closer examination it became apparent that all three                       
states were in fact the same and it was only the analysis                         
techniques that made them appear different.                                                                                                 
The difference in mass between the states observed by the Crystal Barrel 
collaboration and the GAMS collaboration could be reconciled if they both used
the same Breit-Wigner parameterisation. 
To reconcile the data from WA91 required invoking the idea that there could be an interference between
coherently produced states.
A previously known example of this coherent interference is the observation of the $f_0(980)$ in central 
production \cite{pipif0sig}.  In the $K K$ channel the $f_0(980)$ appears as a peak near threshold 
(see for example fig. \ref{fi:swave}c), whereas in the $\pi\pi$ channel, due to interference with the low mass s-wave continuum or $\sigma$, it often 
appears as a dip in the mass spectrum (see fig. \ref{fi:swave}a) \cite{pipif0sig}.  
\par  
In a similar way the WA91 collaboration showed that the X(1450) could be explained as being 
due to the coherent interference of the $f_0(1370)$ and the $f_0(1500)$ \cite{re:new4pi}.
Fig. \ref{fi:swave} shows all the $J^{PC}$ = $0^{++}$ states observed by the
WA102 collaboration.  
These states were observed in the $\pi\pi$ \cite{ref:pipi}\cite{ref:coupledchan}, $KK$ \cite{ref:coupledchan}\cite{ref:KK}, $\eta\eta$ \cite{ref:etaeta},
$\eta\eta^{'}$ \cite{ref:etaetap}, $4\pi$ \cite{ref:4pi} and $\omega\omega$ \cite{ref:omegaomega} final states.
In addition to the low mass $\pi\pi$ continuum or $\sigma$ ($f_0(500)$) these are
$f_0(980)$, $f_0(1370)$, $f_0(1500)$, $f_0(1710)$ and $f_0(2000)$. 
\par
The fact that there are three clearly established states, the $f_0(1370)$, $f_0(1500)$ 
and $f_0(1710)$, in a region where the standard nonet structure would only
require two and that this is the mass region expected for the scalar glueball, meant that it was 
possible that one of these states could be the scalar glueball. 
However, none of them had the decay modes characteristic of a glueball state
and so it was postulated that the observed states are in fact a result of the mixing
of the glueball with the nearby $q \overline q$ states with the same                  
$J^{PC}$ (see \cite{scalars} and references therein). 
Such a mixing would lead to three isoscalar states and the relative $gg$ to $q \overline q$ content of these states would lead to 
a predictable pattern of
decay branching ratios \cite{re:AC,re:FC89,re:CFL}. 
\par
 
The observation of these states in a wide range of decay modes meant that, 
for the first time, a complete data set 
of the two-body decays of these states was available and this was used to extract their 
quark content. The WA102 data, presented in fig. \ref{fi:swave}, produced the following
relative decay rates \cite{scalars}:\\
$ 
\pi \pi : K \overline K : \eta\eta : \eta\eta^\prime : 4\pi
$ \\
$f_0(1370)$:
$ 
1 : 0.46 \pm 0.19 : 0.16 \pm 0.07 : 0.0 : 34.0 ^{+22}_{-9} 
$\\ 
$f_0(1500)$:
$ 
1 \;:\; 0.33 \pm 0.07\; 
:\; 0.18 \pm 0.03\; :\; 0.096 \pm 0.026 \;
:\; 1.36 \pm 0.15
$\\ 
$f_0(1710)$:
$ 
1 : 5.0 \pm 0.7 : 2.4 \pm 0.6 
: <\;0.18\;(90\;\%\;\; CL) : <\;5.4\;(90\;\%\;\; CL) 
$\\ 
\par
These data were used as input to a fit to
investigate 
the glueball-quarkonia content of the $f_0(1370)$, $f_0(1500)$ and 
$f_0(1710)$. 
In the $|G\rangle=|gg\rangle$, $|S\rangle=|s\bar{s}\rangle$, 
$|N\rangle=|u\bar{u}+d\bar{d}\rangle/\sqrt{2}$ basis, 
the physical states $|f_0(1710)\rangle$, $|f_0(1500)\rangle$ and 
$|f_0(1370)\rangle$ were found to be \cite{scalars} 
\[
|f_0(1710)\rangle=0.42|G\rangle+0.89|S\rangle+0.17|N\rangle,
\]
\[
|f_0(1500)\rangle=-0.61|G\rangle+0.37|S\rangle-0.69|N\rangle,
\]
\[
|f_0(1370)\rangle=0.65|G\rangle-0.15|S\rangle-0.73|N\rangle.
\]

\par
\begin{figure}[hb]
\centerline{\includegraphics[width=12cm]{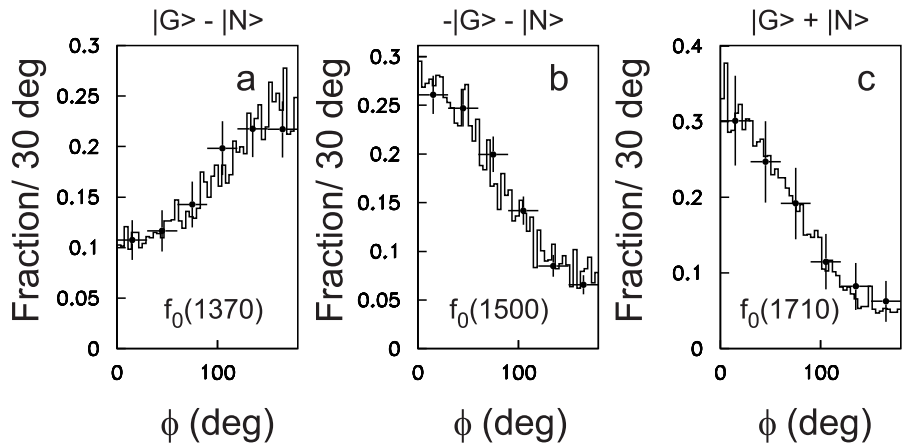}}
\caption{The $\phi$ dependence for a) $f_0(1370)$, b) $f_0(1500)$ and c) $f_0(1710)$ for the data (dots) and the model (histogram).
}
\label{fi:phiangf0}
\end{figure}

The transformation matrix obtained, although not imposed, is close to being unitary indicating the robustness of the solution \cite{scalars}.
The solution shows that the glue content is shared between the three states, it is in phase with the $n \overline n$ content in the $f_0(1500)$ and $f_0(1710)$
and out of phase in the $f_0(1370)$. The $s \overline s$ content is dominantly in the $f_0(1710)$. 
This solution was consistent with the production
of these states 
in $\gamma \gamma$ collisions, ${\it p \bar{p}}$ annihilations and in radiative $J/\psi$ decays \cite{scalars}.
For central production, 
as was discussed in section 2, the cross sections of well established
quarkonia in WA102 suggest that the
production of $s \overline s$ is strongly suppressed 
relative to $n \overline n$. 
The relative 
cross sections for
the three states are
${\it p p} \to {\it pp} + ( f_0(1710):
 f_0(1500) :f_0(1370)) \sim 0.14:1.7:1. $
This would be natural if the production were via the
$n \overline n$ and $gg$ components of DPE.
There is an important qualitative difference for these three states in 
the distribution of the the azimuthal angle $\phi$ between the $p_T$ 
vectors of the two outgoing protons (see fig. \ref{fi:phiangf0}).
The $f_0(1710)$ and $f_0(1500)$ peak as
$\phi \to 0$ whereas the $f_0(1370)$ is more peaked 
as $\phi \to 180^{o}$. While it is possible to explain these observations by postulating that 
the $gg$ and $n \overline n$
components are produced coherently (positive $\mu^{2}$) as $\phi \to 0$ but are out of phase (negative $\mu^{2}$)
as $\phi \to 180$ \cite{scalars}, no such explanation for the observed production of these states as a function of $dP_T$
has been attempted.

\section{The search for the tensor glueball}
\begin{figure}[ht]
\centerline{\includegraphics[width=12cm]{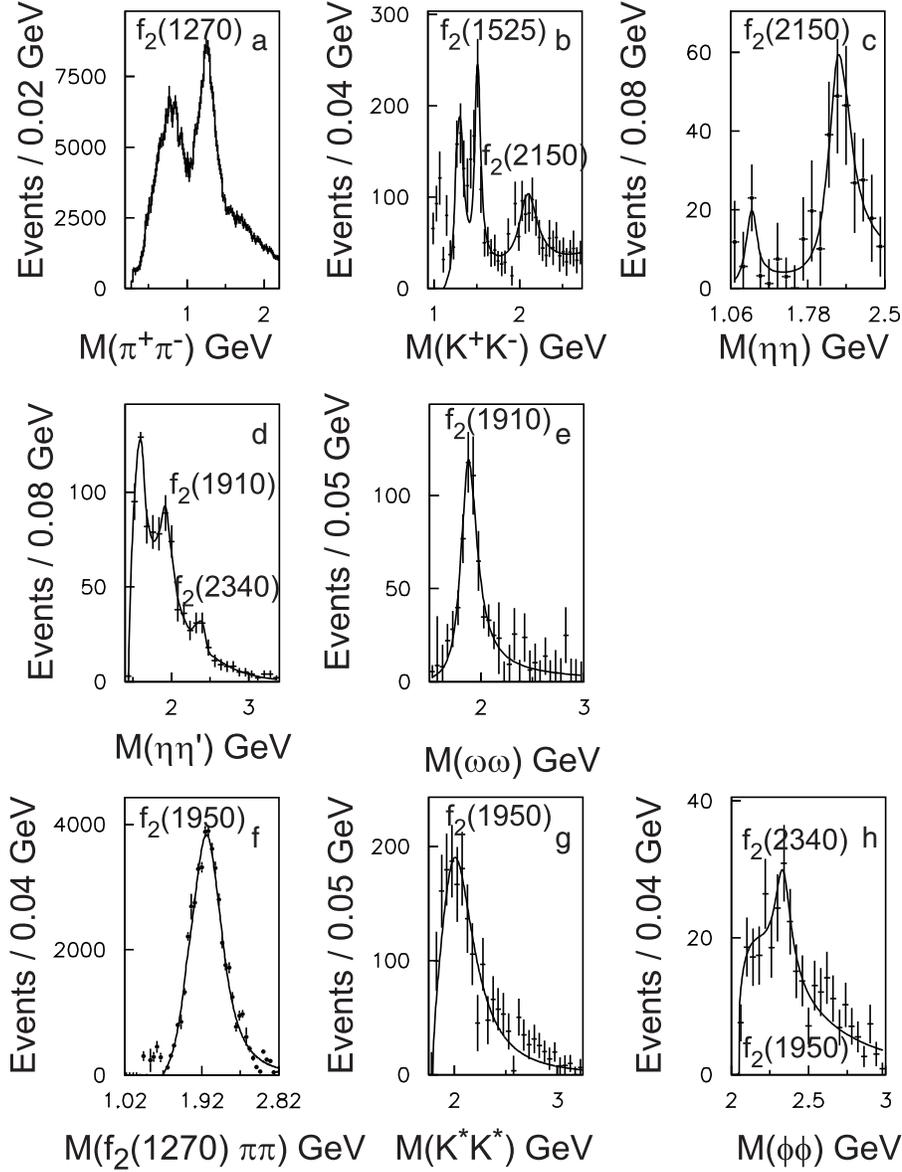}}
\caption{The D-wave contributions to the a) \pipi, b) \kk, c) $\eta\eta$, e) $\omega\omega$, f) $f_2(1270)\pi\pi$, g) $K^{*}K^{*}$ and 
h) $\phi\phi$ mass spectra 
and d) the total $\eta\eta^{'}$ mass spectra.
The location of the $J^{PC}$ = $2^{++}$ states identified are indicated.
}
\label{fi:dwave}
\end{figure}
The tensor sector is much less explored and the level of cross machine analysis is much lower than in the scalar sector.
However, interesting candidates do exist.  The WA76 collaboration reported the observation
of a previously unobserved meson with a mass of 1950 MeV and width of 450 MeV
in the \pipipipi~final state
at a incident beam momentum of 300 GeV \cite{wa764pi}.
A spin parity analysis performed by the WA102 collaboration showed that it has 
$I^{G}(J^{PC})=0^{+}(2^{++})$, with $J_Z=0$ and decays dominantly to $f_2(1270)\pi\pi$ with 
smaller decays to $K^{*}K^{*}$ and $\phi\phi$ \cite{ref:4pi}. 
The observation of this state in DPE combined with the fact that the mass coincides with that 
expected for a state with J=2 on the Pomeron trajectory \cite{re:pomeron_review} led to speculation that this may be the tensor glueball.
However, this was not the only tensor state observed in this mass region. 
Figure \ref{fi:dwave} shows all the states with $I^{G}(J^{PC})=0^{+}(2^{++})$ observed by WA102. 
The isoscalar ground state nonet members, the $f_2(1270)$ and
$f_2(1525)$, are clearly observed in the $\pipi$ and $\kk$ D-wave respectively, 
both are produced with $J_{Z}=0$.  
A state called the $f_2(2150)$ is observed decaying to $\kk$ and $\eta\eta$. 
Another state in the 1.9 GeV region is observed in the $\eta\eta^{'}$ and $\omega\omega$ final states.  The reason that this
state is thought to be distinct from the $f_2(1950)$ is due to the fact that it is produced exclusively with spin projection $J_Z=2$; the only state 
observed in WA102 with $J_Z \ne 0$. It is also narrower with a width of 200 MeV.    
\par
\begin{figure}[ht]
\centerline{\includegraphics[width=12cm]{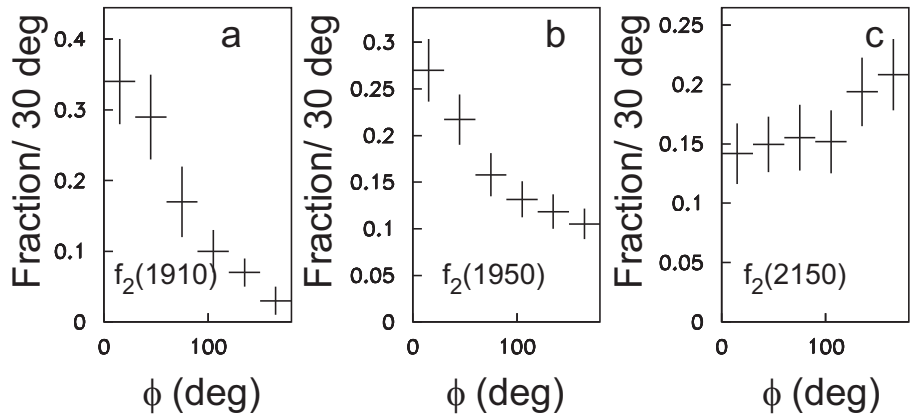}}
\caption{The $\phi$ dependence for a) $f_2(1910)$, b) $f_2(1950)$ and c) $f_2(2150)$.
}
\label{fi:phiangf2}
\end{figure}

As in the scalar sector there appears to be three states where only two would be expected and 
as can be seen from fig.~\ref{fi:phiangf2} the $\phi$ distributions show that 
the $f_2(1910)$ and $f_2(1950)$ peak at $0^o$, while the $f_2(2150)$ peaks near to $180^o$.
Unfortunately there is not enough information on their decay modes to try to extract their relative glue content.  
However, like the $f_0(1500)$ and $f_0(1710)$ in the scalar sector the $f_2(1910)$ and $f_2(1950)$ are produced predominantly at small 
$dP_T$.
While the $f_2(1950)$ is accepted as an established state and appears in the summary tables of the Particle Data Group book \cite{pdg2012} 
the $f_2(1910)$ is not included.
To resolve the tensor glueball further studies are required. 

\section{Conclusions}
\par
Quantum ChromoDynamics (QCD) not only describes how quarks and
antiquarks interact to form the standard $q \overline q$
mesons but also predicts
the existence of glueballs, hybrids and four-quark states.
These states should be produced in 
Double Pomeron Exchange.
Fixed target experiments at the CERN Omega spectrometer have studied centrally produced mesons at
centre-of-mass energies of
$\sqrt{s} = 12.7$, 23.8 and 29~GeV.
This range of energies has allowed the relative strength of the DPE process to be investigated.  
Cuts on the 
$dP_T$ variable select out known $q \bar{q}$ states from
non-$q \overline q$ or glueball candidates. Why this works is still to be understood.
The azimuthal angle $\phi$ has given information on the 
nature of the Pomeron, which is consistent with it transforming like a non-conserved vector current.
Based on the decay rates of the scalar states observed by WA102 and on the hypothesis that
the scalar glueball
mixes with the nearby $q \overline q$ nonet states,
the flavour content of the 
$f_0(1370), f_0(1500)$ and $f_0(1710)$ has been determined.
The solution found 
is also compatible with the relative production 
strengths of the $f_0(1370), f_0(1500)$ and $f_0(1710)$
in $\gamma \gamma$ collisions, ${\it p \bar{p}}$ annihilations
and $J/\psi$ radiative decays.  Tensor candidates exist but further studies are required to resolve this sector.

\end{document}